\documentclass[apj,twocolumn,twocolappendix,floatfix]{openjournal}

\DeclareRobustCommand{\VAN}[3]{#2}
\let\VANthebibliography\thebibliography
\def\thebibliography{\DeclareRobustCommand{\VAN}[3]{##3}\VANthebibliography}

\usepackage{amsmath}	
\usepackage{amssymb}	
\usepackage{booktabs}
\usepackage{lipsum}
\usepackage[T1]{fontenc}
\usepackage{graphicx}	
\usepackage[breaklinks,colorlinks,citecolor=blue,urlcolor=blue,linkcolor=blue]{hyperref}
\usepackage{cleveref}
\usepackage{multirow}
\usepackage{newtxtext,newtxmath}
\usepackage{orcidlink}
\usepackage{pifont}
\usepackage[nobottomtitles]{titlesec}
\usepackage{natbib}
\usepackage{xparse}
\usepackage{ifthen}
\titleformat{\section}{\filcenter\MakeUppercase}{\thesection.}{0.5em}{}

\usepackage{savesym}
\savesymbol{tablenum}
\usepackage{siunitx}
\restoresymbol{SIX}{tablenum}
\usepackage{needspace}
\usepackage{overpic}
\usepackage{xcolor}
\usepackage{lipsum}

\begin{document}

\title{Unraveling the Nature of the Nuclear Transient AT2020adpi}

\author{\vspace{-1.3cm}
Paarmita Pandey\,\orcidlink{0009-0003-6803-2420}$^{1,2}$,
Jason~T.~Hinkle\,\orcidlink{0000-0001-9668-2920}$^{3}$,
Christopher~S.~Kochanek\,\orcidlink{0000-0002-1790-3148}$^{1,2}$,
Michael~A.~Tucker\,\orcidlink{0000-0002-2471-8442}$^{1,2}$,
Mark~T.~Reynolds\,\orcidlink{0000-0003-1621-9392}$^{1,4}$,\\
Jack~M.~M.~Neustadt\,\orcidlink{0000-0001-7351-2531}$^{5}$,
Todd~A.~Thompson\,\orcidlink{0000-0003-2377-9574}$^{1,2,6}$,
Katie~Auchettl\,\orcidlink{0000-0002-4449-9152}$^{7,8}$,
Benjamin.~J.~Shappee\,\orcidlink{0000-0003-4631-1149}$^{3}$,
Aaron Do\,
\orcidlink{0000-0003-3429-7845}$^{9}$,\\
Dhvanil~D.~Desai\,
\orcidlink{0000-0002-2164-859X}$^{3}$,
W.~B.~Hoogendam\,\orcidlink{0000-0003-3953-9532}$^{3,a}$,
C.~Ashall\,\orcidlink{0000-0002-5221-7557}$^{3}$,
Thomas~B.~Lowe\,\orcidlink{0000-0002-9438-3617}$^{3}$,
Melissa~Shahbandeh\,\orcidlink{0000-0002-9301-5302}$^{5,10}$
and
Anna~V.~Payne\,\orcidlink{0000-0003-3490-3243}$^{10}$\\
}

\affiliation{
$^{1}$Department of Astronomy, The Ohio State University, 140 W. 18th Ave., Columbus, OH 43210, USA\\
$^{2}$Center for Cosmology and Astroparticle Physics, The Ohio State University, 191 W. Woodruff Ave., Columbus, OH 43210, USA\\
$^{3}$Institute for Astronomy, University of Hawai'i at Manoa, 2680 Woodlawn Dr., Hawai'i, HI 96822, USA\\
$^{4}$Department of Astronomy, University of Michigan, 1085 S. University Ave., Ann Arbor, MI 48109, USA\\
$^{5}$Department of Physics and Astronomy, Bloomberg Center, Johns Hopkins University, Baltimore, MD 21218, USA\\
$^{6}$Department of Physics, The Ohio State University, 191 W. Woodruff Ave, Columbus, OH 43210, USA\\
$^{7}$School of Physics, The University of Melbourne, Parkville, VIC 3010, Australia.\\
$^{8}$Department of Astronomy and Astrophysics, University of California, Santa Cruz, CA 95064, USA\\
$^{9}$Institute of Astronomy and Kavli Institute for Cosmology, University of Cambridge, Madingley Road, Cambridge CB3 0HA, UK\\
$^{10}$Space Telescope Science Institute, 3700 San Martin Drive, Baltimore, MD 21218, USA\\
$^{a}$NSF Graduate Research Fellow
}

\begin{abstract}
Transient events associated with supermassive black holes provide rare opportunities to study accretion and the environments of supermassive black holes. We present a multiwavelength study of AT2020adpi (ZTF20acvfraq), a luminous optical/UV transient in the nucleus of the galaxy WISEA J231853.77$-$103505.6 ($z=0.26$) that exhibits the properties of an ambiguous nuclear transient. Near peak, its spectral energy distribution is well described by a power law ($\lambda L_\lambda \propto \lambda^{-\alpha}$, $\alpha = 0.44 \pm 0.04$), with a maximum $g$-band luminosity of $(3.6 \pm 0.6)\times10^{44}$ erg s$^{-1}$, which is consistent with luminous AGN flares. We detect a strong mid-infrared flare ($L_\mathrm{peak}^{\mathrm{MIR}} = (2.3 \pm 0.05)\times10^{44}$ erg s$^{-1}$) delayed by $\sim$210 rest-frame days, indicating a hot dust echo from material at $\sim$0.2 pc. The optical and near-infrared spectra show broad H, \ion{He}{1}, [O III] lines, as well as narrow \ion{Fe}{2}, and prominent \ion{Mg}{2}, which is a combination not typical of TDEs. Taken together, these features suggest AT2020adpi is an ambiguous nuclear transient, where an accretion episode was triggered by stellar disruption of an accretion disk or instabilities within an active nucleus. This source demonstrates the need for careful multiwavelength analysis to distinguish between extreme AGN variability and TDEs.
\end{abstract}

\maketitle

\section{Introduction}

Active galactic nuclei (AGNs) are known to vary in both their photometric brightness and spectroscopic features. This variability is primarily due to changes in the rate at which material is accreted onto the central supermassive black hole (SMBH), although variable obscuration may also play a role. While AGN variability has been studied for decades (e.g., \citealt{andrillat68,tohline76,oknyanskij78, Padovani2017, Paolo2025}), recent observations have revealed a broader range of events that differ significantly from the typical, low-amplitude, stochastic fluctuations commonly seen in AGN (e.g., \citealt{macleod12, komossa2024}).

Among the most intriguing of these are changing-look (CL) AGN (e.g., \citealt{bianchi05,shappee14,macleod16,trakhtenbrot19, Mcloed2019, sheng2017, Ricci2023, Guo2025}), where broad emission lines appear or disappear on timescales ranging from months to years. Similarly, rapid turn-on events, including changing-look low-ionization nuclear emission regions (LINERs; e.g., \citealt{gezari17,yan19,frederick19}), involve a previously inactive or weakly active galactic nucleus suddenly transitioning to an AGN-like state. In addition, other transient phenomena associated with SMBHs have been identified that do not easily fit into existing classification schemes (e.g., \citealt{kankare17,tadhunter17,gromadzki19,trakhtenbrot19-17cv, sheng2020}).

Another important class of SMBH-driven transients is Tidal Disruption Events (TDEs). These occur when a star passes too close to a SMBH and is torn apart by its tidal forces \citep{rees88,phinney89,evans89}. A fraction of the disrupted stellar material is then accreted onto the black hole, generating a luminous transient that emits in the optical, ultraviolet, and, sometimes, radio and X-rays, gradually fading over time. The characteristics of these events depend on several factors: the orbital parameters of the disrupted star \citep{guillochon13,dai18}, the physical properties of both the star and the black hole \citep{guillochon13,kochanek16}, and the effects of radiative feedback from accretion \citep{gaskell14,strubbe15,roth16,roth18}.

TDEs typically radiate on the order of $10^{51}$erg in observable bands \citep{holoien14-14ae,auchettl17, Mockler_2021}, corresponding to the conversion of less than $0.01M_\odot$ of mass into radiation under the assumption of a $10\%$ accretion efficiency ($\eta = 0.1$). This suggests that most bound debris is expelled rather than accreted, or that the accretion efficiency is lower than expected. However, when the emission is integrated over the years-long fading phase beyond the initial months of peak activity, the total radiated energy may approach that expected from the accretion of up to $\sim 0.1~M_\odot$ \citep{vanvelzen19}.

TDEs and AGN differ markedly in their emission properties across wavelengths. Optically selected TDEs are dominated by a strong UV/optical blackbody spectral energy distribution (SED) with nearly constant temperatures of $20,000 - 50,000$ K \citep{gezari12b, holoien14a, vanvelzen21}. The existence of UV flux beyond the blackbody has been known for a long time due to the presence of recombination lines \citep{gezari12, holoien16-14li, hung17}, and the blackbody temperature measurements from TDEs are regularly underestimated \citep{Arcavi_2022, guolo2025}. The UV/optical SEDs of AGNs are usually best fit with power laws, $\lambda L_\lambda \propto \lambda^{-\alpha}$ \citep[e.g.,][]{vandenberk01, temple2023}, with $1<\alpha<2$ \citep[e.g.,][]{koratkur99,vandenberk01}. Some TDEs emit soft X-rays (kT$ \sim 30 - 60$ eV) \citep{holoien16a, holoien16b, hinkle21b, wevers20}, and are typically much softer than the X-ray emission from AGN, which extends beyond 10 keV due to Comptonization \citep{ricci17, auchettl17, auchettl18}. TDEs generally show a rapid, luminous flare with a smooth peak and monotonic decline \citep{holoien14a, holoien19b, nicholl16, auchettl18, hinkle21b, hammerstein2023}. In contrast, AGNs exhibit stochastic variability and occasional rebrightening in flares \citep{macleod16}. Moreover, TDEs typically occur on lower-mass SMBHs ($\leq 10^{7}~M_\odot$) \citep{Mockler2023}, since main-sequence stars approaching more massive black holes would pass through the event horizon without disruption \citep{Hills1975, rees88}.

Spectroscopically, TDEs feature blue continua and broad H and/or He emission lines, often with full width at half maximum (FWHM) $\geq10,000$ km/s, sometimes accompanied by metal lines or Bowen fluorescence \citep{arcavi14, holoien16a, holoien16b, blagorodnova17, leloudas19, hammerstein2023}. In contrast, AGN spectra typically display emission lines of width $\sim5000$ km/s, including prominent Balmer and forbidden lines like [OIII] and [NII] \citep{vandenberk01, frederick20, sheng2020}. Their UV spectra also differ significantly. In particular, AGNs show strong \ion{Mg}{2}  ($\lambda ~2798$\AA) emission, which is absent in TDEs \citep{brown18, hung20, vanvelzen21, Yao2023}. The line evolution with respect to luminosity also differs; TDEs generally exhibit a positive correlation between line luminosity and line width, leading to narrower emission lines as the TDE fades \citep{holoien19a, hinkle21a, Charalampopoulos_2022}, whereas AGNs show the inverse trend, where the Balmer lines broaden as AGN fades \citep{peterson04, denney09, Shen2012}.

The advent of wide-field, non-targeted sky surveys such as the All-Sky Automated Survey for Supernovae \citep[ASAS-SN;][]{shappee14, kochanek17}, the Asteroid Terrestrial Impact Last Alert System \citep[ATLAS;][]{tonry18}, the Zwicky Transient Facility \citep[ZTF;][]{bellm19}, the Young Supernova Experiment \citep{Jones_2021}, and the Panoramic Survey Telescope and Rapid Response System \citep[Pan-STARRS;][]{chambers16} has significantly expanded the range of transient phenomena observed in galactic nuclei. This has led to increased detection of nuclear transients such as CL-AGN and TDEs, and transients whose physical origins remain ambiguous \citep[e.g.][]{ kankare17,neustadt20,hinkle22, Hinkle_science_25}. These events have been named ambiguous nuclear transients (ANTs) since they do not fit into the known categories of standard AGN variability, TDEs, or supernovae (SNe; \citealt{Wiseman25, Hinkle2024}). The ANTs often display peculiar features and involve substantial changes in previously steady accretion systems. These behaviors suggest complex interactions, such as an SNe or TDE disrupting an existing AGN disk. Some recent examples include PS1-10adi \citep{kankare17}, ASASSN-18jd \citep{neustadt20}, ASASSN-18el \citep{trakhtenbrot19b, ricci20},  ASASSN-17jz \citep{Holoien_2022}, AT2018dyk \citep{frederick19}, ASASSN-20hx \citep{Hinkle_2022}, AT2021loi \citep{Makrygianni_2023}, and AT2021lwx \citep{Wiseman2023, Subrayan_2023}. These nuclear outbursts are most likely powered by accretion onto SMBHs, potentially offering new insight into black hole accretion processes, especially in galaxies that are otherwise dormant. A distinct subclass of ANTs, known as extreme nuclear transients (ENTs), represents the most energetic transients currently observed. These events may occur due to the tidal disruption of massive stars ($3$–$10~M_\odot$) \citep{Hinkle_science_25, Graham_2025}.

In this paper, we present observations of the ANT AT2020adpi (ZTF20acvfraq/ATLAS20bjzp/Gaia21aid, RA=$349.72405$, Dec=$-10.58495$) at $z=0.26$ \citep{ZTF_report21}. This transient was first analyzed in \cite{Wiseman25} as a part of a study of 11 ANTs. Further details and comparison with their analysis are detailed in Section \ref{discussion}. Section \ref{obs} describes the photometric and spectroscopic observations of the transient. In Section \ref{results}, we examine the properties of the host galaxy, derive the UV/optical SED, and other key characteristics of the transient. Finally, in Section \ref{discussion}, we discuss the nature of AT2020adpi within the broader context of TDEs, standard AGN variability, and the emerging class of ANTs.  We adopt a luminosity distance of $D_{L} = 1360.4$~Mpc for a flat universe with $h = 0.696$, $\Omega_{\rm M} = 0.286$, and $\Omega_\Lambda = 0.714$ \citep{wright06}. We correct the photometry for a Galactic extinction of $A_V = 0.074$ mag \citep{Extinction2011}.

\begin{figure*}[t]
    \begin{center}
\includegraphics[width=\textwidth]{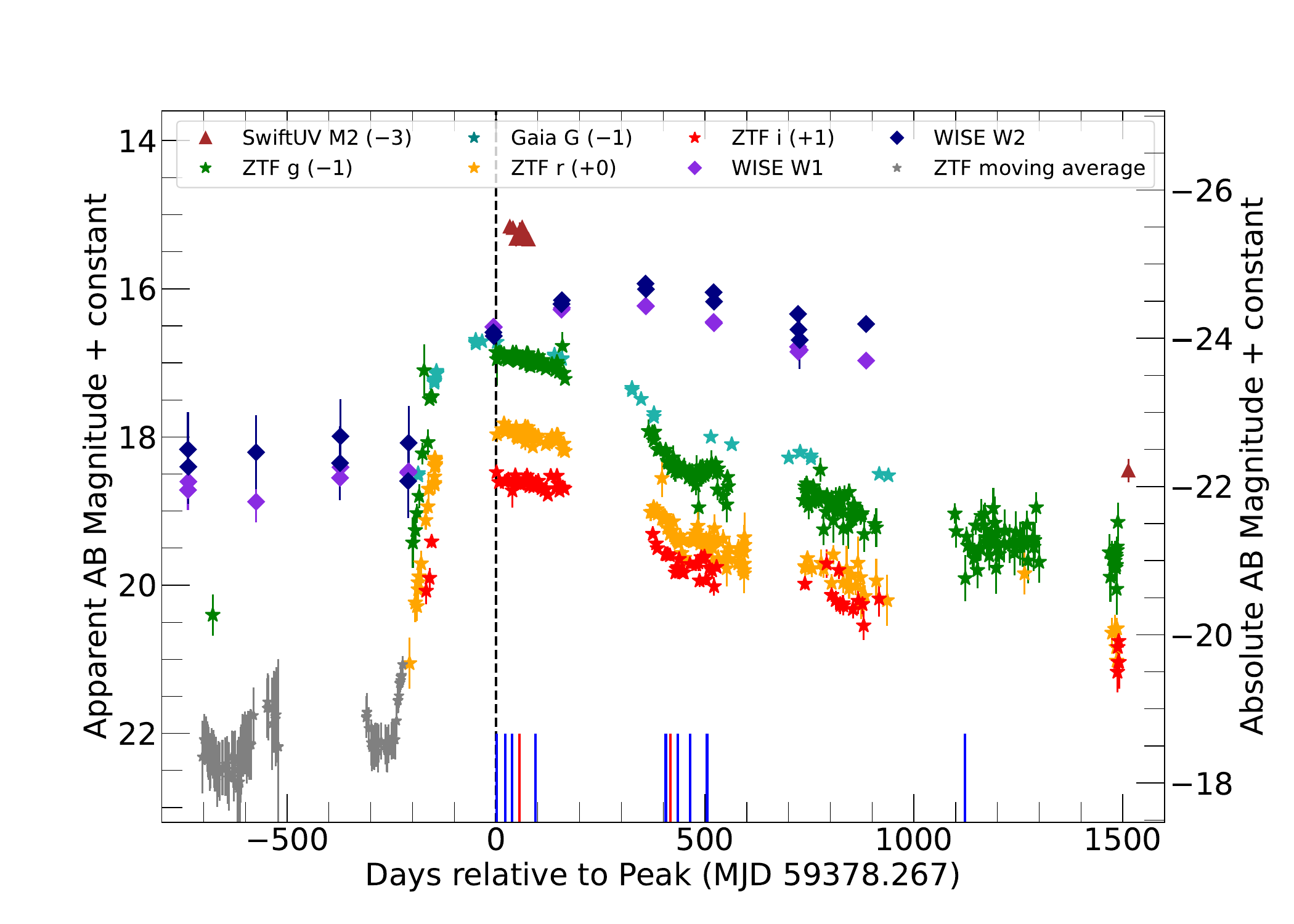}
       \caption{Host-subtracted UV, optical, and IR light curves of AT2020adpi, showing the Swift UVM2 band, ZTF $g$,$r$, $i$ bands, Gaia $G$ band, and WISE $W1$ and $W2$ bands. The photometry spans from roughly 800 days before peak (MJD = 59378.267) to roughly 1500 days after peak in observer-frame days. The gray ZTF points are the 100-day moving average of the pre-event ZTF fluxes. The blue (red) bars along the time axis show the epochs of optical (near-IR) spectroscopic follow-up. The black line is the adopted time reference. All data are corrected for Galactic extinction and are in the AB magnitude system, including the WISE data. The light curves have been offset for visual clarity. } 
       \label{fig:all_lc}
    \end{center}
\end{figure*}

\begin{figure*}
    \begin{center}
\includegraphics[width=18cm, height=14cm]{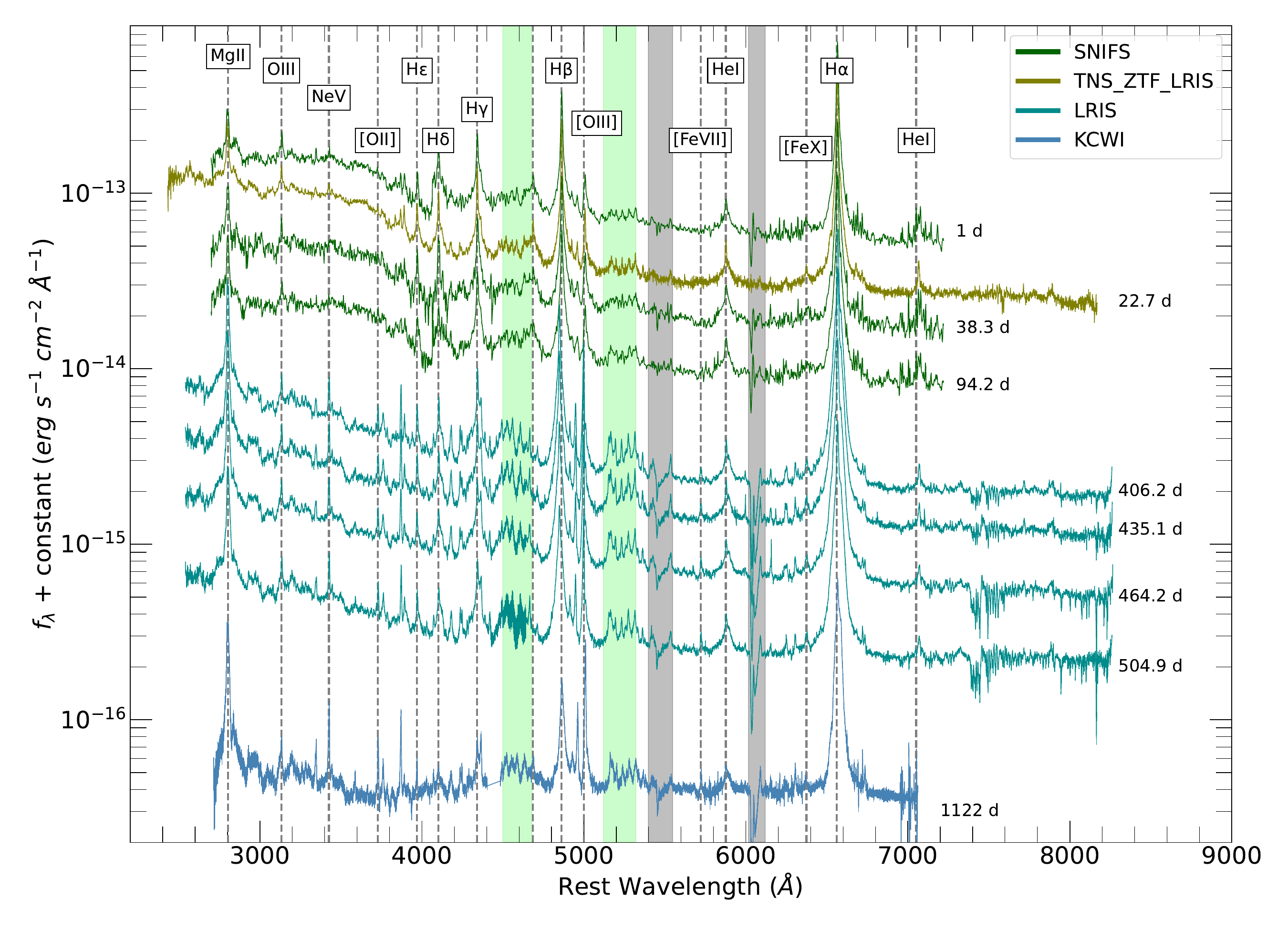}
\vspace{-3mm}
       \caption{SNIFS, LRIS, and KCWI optical spectra of AT2020adpi, where the time after peak in the observed frame is to the right of each spectrum. The spectra span from almost 1 day after peak UV/optical emission (top) until 1122 days after peak (bottom). Prominent emission lines are labeled. The green shaded region highlights the \ion{Fe}{2} emission complex. The gray shaded regions show the telluric bands. The spectra are offset for visibility.} 
       \label{fig:kcwi}
    \end{center}
\end{figure*}

\section{Observations and Survey Data}\label{obs}
This section summarizes the archival data available for the host galaxy and our new photometry and spectroscopy of AT2020adpi. All the data were corrected for Galactic extinction.

\subsection{Photometry}\label{sec:photometry}
We retrieved the Zwicky Transient Facility \citep[ZTF;][]{Graham2019, Bellm2019} optical light curve of AT2020adpi spanning the period from 17 March 2018 (MJD 58194) to 21 July 2025 (MJD 60877). We used the IPAC ZTF forced photometry server $g$, $r$, and $i$ band data \citep{masci2023newforcedphotometryservice}, following the recommended procedures\footnote{\url{https://irsa.ipac.caltech.edu/data/ZTF/docs/ztf_forced_photometry.pdf}}. To assess any faint pre-event activity in the light curve, we computed 100-day moving averages of the difference-image fluxes relative to the reference image, to search for any low-level variability. We also used the publicly available optical photometry from Gaia Data Release 3 \citep[DR3;][]{Gaia_21}. We used W1 and W2 mid-infrared (MIR) data from the Wide-field Infrared Survey Explorer \citep[WISE;][]{Mainzer2011} for our analysis. We retrieved light curves using the method described by \cite{WISE-pull-lc}. We group the data into 24-hour bins and use the median magnitude within each bin. The uncertainty is derived from the photometric error estimates.

During the period from MJD 59411 $-$ 59455 ($\approx 30-80$ days post peak), the Neils Gehrels Swift Observatory observed the transient using all six ultraviolet and optical (UVOT) \citep{poole08} filters: $V$ (5468~\AA), $B$ (4392~\AA), $U$ (3465~\AA), $UVW1$ (2600~\AA), $UVM2$ (2246~\AA), and $UVW2$ (1928~\AA) for $\sim2500$ s. We also requested a Swift target of opportunity observations with the UVM2 filter for the UVOT and XRT for 4 kiloseconds in July 2025, approximately 1500 days post-peak brightness. For each epoch, two exposures were obtained for each filter. These were combined into single images using the \texttt{HEASoft} utility {\tt uvotimsum}. Source counts were then extracted using {\tt uvotsource} with a circular aperture of radius 5\farcs{0} centered on the transient. Background counts were measured from a nearby, source-free region with a radius of $\sim$50\farcs{0}. The resulting count rates were converted into physical fluxes and magnitudes using HEASoft \textit{version 6.34} and the CALDB release {\tt 2024-08-12\_V6.34} \citep{poole08, breeveld10}. 

Figure~\ref{fig:all_lc} shows the UV/optical/mid-IR light curves of AT2020adpi, and Table~\ref{tab:phot} provides the photometric data. We only show the Swift UVM2 band of the UVOT filters.  We chose MJD 59378.267 as the reference epoch for our light curve and spectral parameters; the reason for choosing this epoch is described in Section~\ref{LC}. The gray data points are 100-day moving averages of the ZTF fluxes. They suggest a small jump in flux starting roughly about two years before peak, possibly with a small flare prior to the transient.

\begin{deluxetable}{cccc}
\tablewidth{240pt}
\tabletypesize{\footnotesize}
\tablecaption{Extinction Corrected Photometry of AT2020adpi}
\tablehead{
\colhead{MJD} &
\colhead{Filter} &
\colhead{Magnitude (AB)} & 
\colhead{Uncertainty} }
\startdata
59411.286 &	swift UVW2	& 18.334 & 0.061\\ 
59418.733 & swift UVW2	& 18.432 & 0.092\\
59425.634 & swift UVW2	& 18.273 & 0.102\\
59434.917 & swift UVW2	& 18.334 & 0.102\\
\ldots  & \ldots & \ldots & \ldots  \\
60101.993 & WISE W2	& 16.551 & 0.143 \\
60104.993 &	WISE W2	& 16.692 & 0.388 \\
60263.993 &	WISE W2	& 16.4755 & 0.079 
\enddata 
\tablecomments{Extinction-corrected magnitudes and associated uncertainties for the photometry of AT2020adpi. All magnitudes are reported in the AB magnitude system. Only a portion of the table is shown here to illustrate the format; the complete dataset is available online as an ancillary file.} 
\label{tab:phot} 
\end{deluxetable}

The Swift XRT observed the transient simultaneously with the UVOT. No X-ray emission was detected either near peak (total exposure of 9.4 ks), at late times (4.0 ks), or in the combined 13.4 ks dataset. We derive $3\sigma$ upper limits on the X-ray flux incorporating the first set of observations near peak (9.4 ks), 1500 days post peak (4 ks), and the total time of 13.4 ks, assuming a power-law model $dN/dE \propto E^{-\gamma}$, with $\gamma = 1.8$ typical for AGN spectra, and a hydrogen column density of $N_{\mathrm{H}} = 0.024 \times 10^{22}~\mathrm{cm}^{-2}$ \citep{1990Dickey, 2005Kalberla, HI42016}. 
For the near-peak observations, we get a $3\sigma$ flux limit of $f_{X} < 6.9 \times 10^{-14}~\mathrm{erg~s^{-1}~cm^{-2}}$ in the $0.5$--$10.0~\mathrm{keV}$ band, corresponding to $L_X \leq 1.4\times10^{43}~\rm erg~s^{-1}$. The most recent observations give us a flux limit of  $f_{X} < 2.0 \times 10^{-12}~\mathrm{erg~s^{-1}~cm^{-2}}$ ($0.5$--$10.0~\mathrm{keV}$), and $L_X \leq 4.2\times10^{44}~\rm erg~s^{-1}$. Overall, for the combined set of observations, we have $f_{X} < 6.3 \times 10^{-14}~\mathrm{erg~s^{-1}~cm^{-2}}$ in the $0.5$--$10.0~\mathrm{keV}$ band, corresponding to $L_X \leq 1.3\times10^{43}~\rm erg~s^{-1}$.


\begin{figure*}[t]
    \begin{center}
\includegraphics[width=13cm, height=7cm]{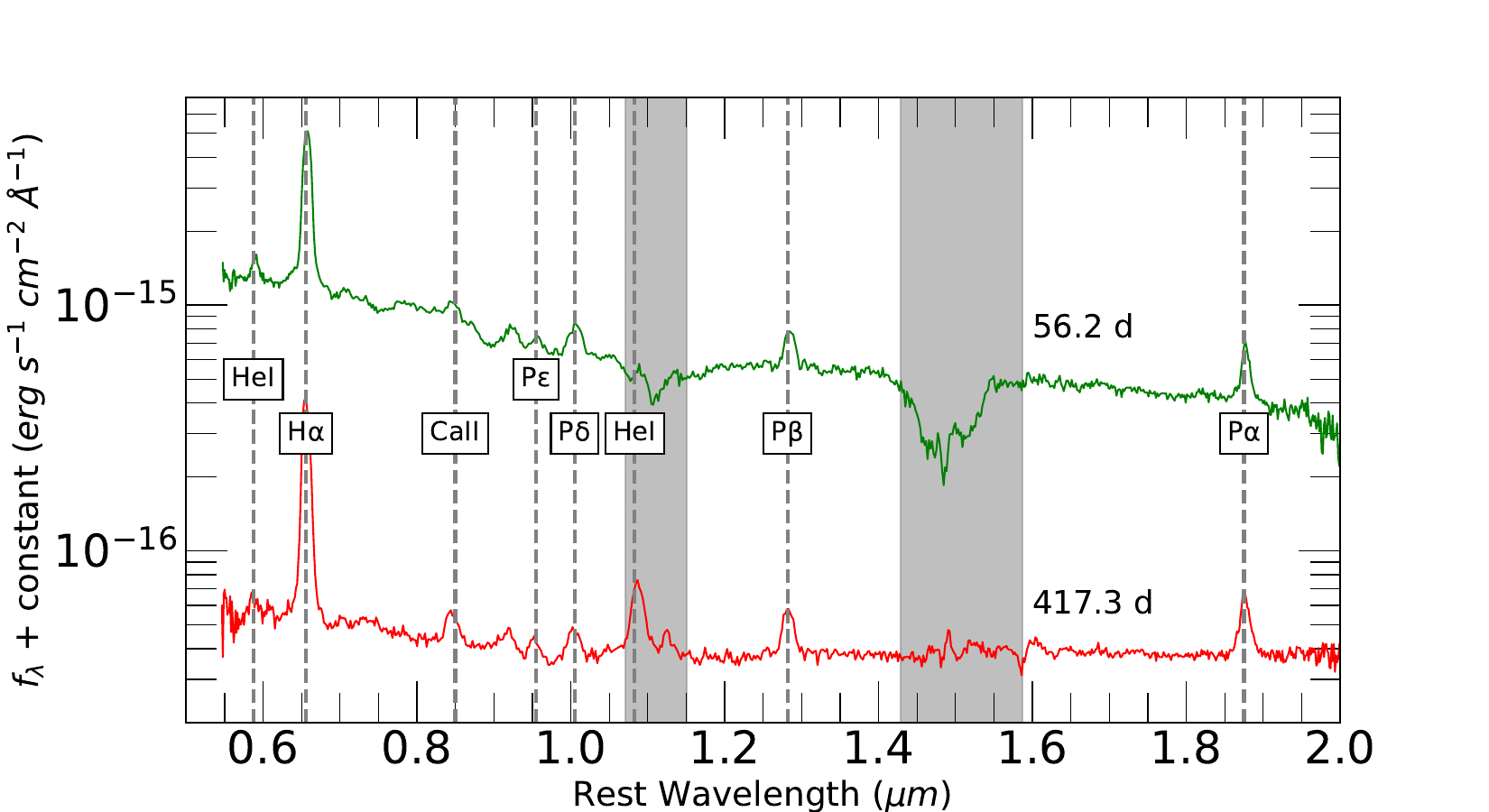}

       \caption{SpeX near-IR Spectra of AT2020adpi, where the time after peak in the observed frame is given above the spectra. Prominent emission lines are labeled. The vertical gray bands mark atmospheric telluric features. Due to an overexposed calibration standard on the night of the first observation, the spectrum taken at 56 days post-peak (green) was re-reduced using a standard from a different night at a similar airmass. This improved the overall spectral shape, but left large residuals in the telluric bands. } 
       \label{fig:spex}
    \end{center}
\end{figure*}
\subsection{Spectroscopic Observations}

We obtained spectroscopic observations of AT2020adpi from several sources. The three spectra from the SuperNova Integral Field Spectrograph \citep[SNIFS;][]{lantz04} on the 88-inch University of Hawai'i telescope (UH88) were obtained near peak and between $30-100$ days post peak. We obtained spectra from the Keck Low-Resolution Imaging Spectrometer \citep[LRIS;][]{oke95} on the 10-m Keck I telescope more than a year after the optical peak. Finally, we obtained a spectrum of AT2020adpi with KCWI on Keck in July 2024 (MJD 60500) when it had nearly faded to the pre-flare level. We also used the TNS spectrum from ZTF taken $\sim 20$ days after the peak \citep{2021TNS}.

The optical spectra from SNIFS were calibrated and reduced using the SCAT pipeline \citep{Tucker_2022}. LRIS spectra were reduced with {\tt PypeIt} \citep{pypeit:joss_arXiv, pypeit:joss_pub, pypeit:zenodo}.
KCWI spectra were reduced with KCWI DRP\footnote{\url{https://kcwi-drp.readthedocs.io/en/latest/index.html}}. An initial flux calibration was performed using spectrophotometric standard stars observed on the same nights as the science targets. Figure~\ref{fig:kcwi} shows the optical spectroscopic evolution of AT2020adpi spanning from the peak UV/optical emission time to $\sim 1100$ days post-peak in the observer's frame. 

In addition, we acquired two near-infrared (NIR) spectra of AT2020adpi using the SpeX \citep{rayner03} on the NASA Infrared Telescope Facility (IRTF). The observations were taken in prism mode with $R\sim80$ since the source was faint for IRTF. These NIR spectra were reduced using {\tt SpeXtool} \citep{Cushing2004}. Due to an overexposed calibration standard on the night of observation, the spectrum taken 56 days post-peak was re-reduced using a standard from a different night at a similar airmass. This improved the overall spectral shape, but the residuals in the telluric bands are large. Figure~\ref{fig:spex} shows the spectroscopic evolution of AT2020adpi in the near IR.  

\begin{figure*}[t]
    \begin{center}
\includegraphics[width=16cm, height=12cm]{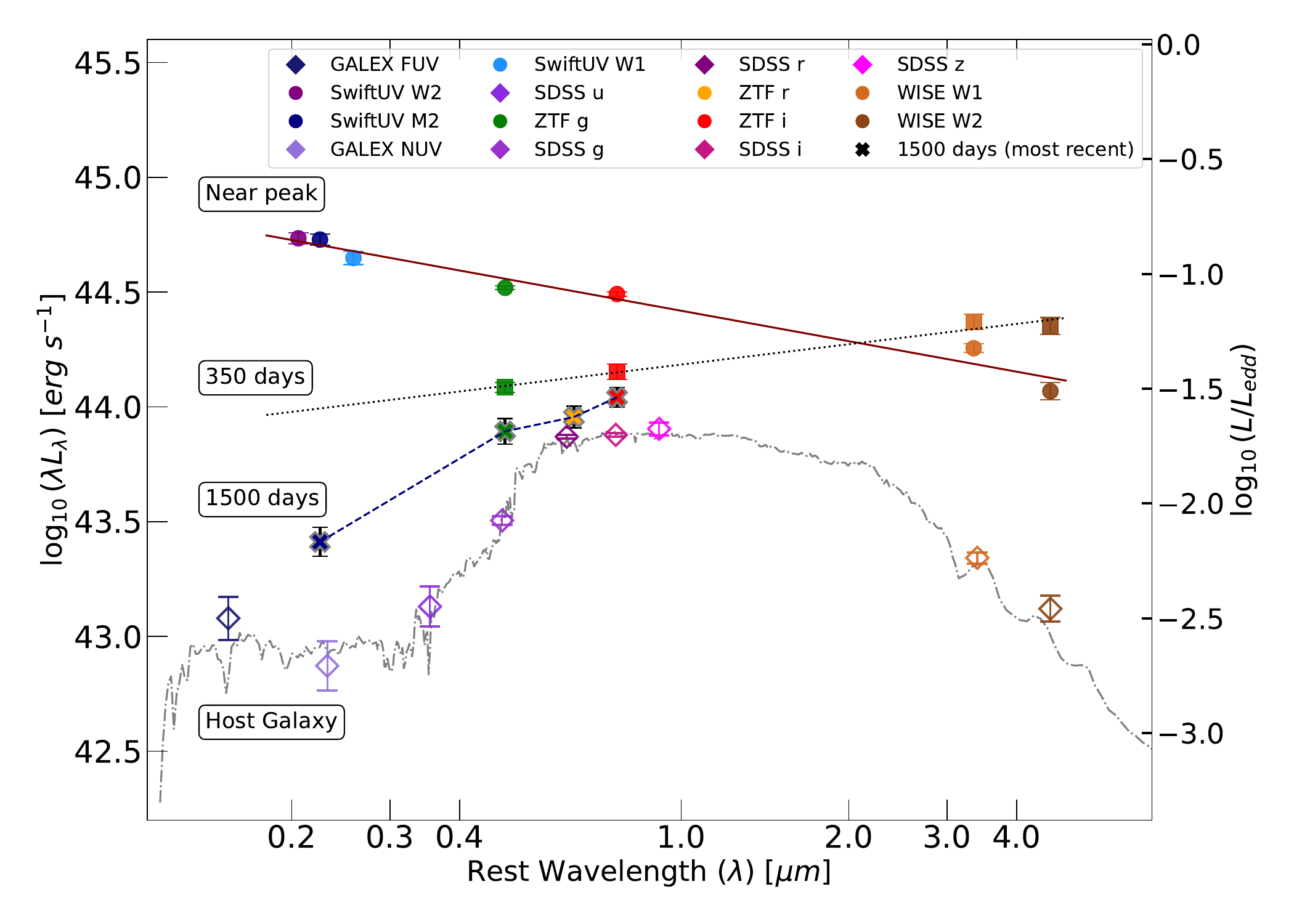}
       \caption{ SED of AT2020adpi near peak, 350 days and 1500 days post-peak. The open markers show the SED of the host galaxy. The FAST SED model is shown for the host and power-law fits $\lambda L_\lambda \propto \lambda^{-\alpha}$ for AT2020adpi, with $ \alpha=0.44\pm0.04$ near peak and $ \alpha=-0.29\pm0.05$ at 350 days post peak. The Eddington luminosity is also indicated on the secondary y-axis, showing that the peak luminosity reaches approximately $15\%$ of the Eddington limit.  } 
       \label{fig:SED_fit}
    \end{center}
\end{figure*}

\section{Results}\label{results}

In this section, we summarize the results obtained from the host galaxy analysis, optical spectra, SED models of AT2020adpi, and the light curve. 

\subsection{Host Galaxy} \label{hostgalaxy}
To characterize the properties of the host galaxy, we fit the pre-transient SED using Fitting and Assessment of Synthetic Templates \citep[\textsc{Fast++};][]{kriek09} to model the archival photometric measurements. We modeled the data from GALEX, SDSS, and WISE data, summarized in Table~\ref{tab:photometry}. 
We used a \citet{cardelli89} extinction law with a total-to-selective extinction ratio of $R_{V} = 3.1$, a Salpeter initial mass function \citep{salpeter55}, and an exponentially declining star formation history (SFH) using \citet{bruzual03} models fixed at solar metallicity. The results are presented in Table~\ref{tab:fast_results}. The host galaxy has a stellar mass of $\log(M_*/M_\odot) = 10.36^{+0.10}_{-0.07}$ and a star formation rate (SFR) of $\log(\mathrm{SFR}/M_\odot\,\mathrm{yr}^{-1}) = -0.27^{+0.15}_{-0.13}$. Figure~\ref{fig:SED_fit} shows the host SED with the FAST model. The SED is consistent with no current star formation due to the weak UV emission. In the model, the last burst of star-formation occurred about $\sim 1$ Gyr ago. Given the stellar population age and lack of current star formation, we conclude that the host is likely a post-starburst galaxy. Unfortunately, there are no pre-event spectra of the host. 
To constrain black hole mass, we first derived the bulge mass from the total stellar mass of the host galaxy using the empirical bulge-to-total scaling relations of \citet{mendel14}. This yielded a bulge mass of $\log(M_{\mathrm{bulge}}/M_\odot) = 10.16$. We then used the \citet{mcconnell13} $M_{\mathrm{BH}}$–$M_{\mathrm{bulge}}$ relation to infer a central black hole mass of $M_{BH} \approx 10^{7.6}~M_\odot $.

\begin{table}
\centering
\caption{Photometric Measurements of Host Galaxy WISEA J231853.77$-$103505.6}
\label{tab:photometry}
\begin{tabular}{lcc}
\toprule
Telescope and Filter & Magnitude & Magnitude Error \\
\midrule
GALEX (AB) FUV   & 22.78 & 0.24 \\
GALEX (AB) NUV   & 22.86 & 0.27 \\
SDSS (AB) $u$    & 21.75 & 0.22 \\
SDSS (AB) $g$    & 20.35 & 0.04 \\
SDSS (AB) $r$    & 19.29 & 0.02 \\
SDSS (AB) $i$    & 19.05 & 0.02 \\
SDSS (AB) $z$    & 18.79 & 0.07 \\
WISE (Vega) W1   & 16.08 & 0.06 \\
WISE (Vega) W2   & 15.67 & 0.14 \\
WISE (Vega) W3   & 11.64 & 0.34 \\
\bottomrule
\end{tabular}
\end{table}

\begin{deluxetable}{lc}
\tablecaption{Stellar Population Parameters from FAST++ \label{tab:fast_results}}
\tablecolumns{2}
\tablewidth{240pt}
\tablehead{
\colhead{Parameter} & \colhead{Value}
}
\startdata
Star formation timescale (log$(\frac{\tau}{\rm yr})$) & $8.40^{+0.06}_{-0.21}$ \\
Stellar population age (log$(\frac{\rm age}{\rm yr})$) & $9.15^{+0.05}_{-0.17}$  \\
Dust attenuation $A_V$ (mag) & $0.08^{+0.12}_{-0.00}$  \\
Stellar mass (log $M_\odot$) & $10.36^{+0.10}_{-0.07}$  \\
Star formation rate (log$(\frac{M_\odot}{\rm yr})$) & $-0.27^{+0.15}_{-0.13}$  \\
Specific star formation rate (log yr$^{-1}$) & $-10.64^{+0.19}_{-0.17}$ \\
 Black hole mass: $\log(\frac{M_{BH}}{M_\odot})$ & 7.6  
\enddata
\label{tab:host_galaxy} 
\tablecomments{Parameters derived using FAST++ with \cite{bruzual03} templates, a Salpeter IMF, an exponentially declining SFH, and total-to-selective extinction ratio $R_v = 3.1$. The uncertainties are 68$\%$ confidence intervals. The black hole mass has been derived using a scaling relation.}
\end{deluxetable}

We also examined the light curve of the host galaxy for variability using archival data from the Catalina Real-Time Transient Survey \citep[CRTS;][]{drake09}, covering the period from MJD 53710 (6 December 2005) to MJD 55398.8 (21 July 2010). No significant variability on timescales of 100 days was observed during this interval. However, the ZTF data suggest there was a small flare two years prior to the transient based on the 100-day moving average of the fluxes (see Fig~\ref{fig:all_lc}).


\subsection{Spectral Energy Distribution}
\label{sec:spectral_energy_dist}
The SED can help us distinguish the origin of the transient. In particular, AGN have power-law UV/optical SEDs while TDEs are typically well modeled by a $\sim30,000$~K blackbody. Figure \ref{fig:SED_fit} shows the SED of AT2020adpi around the optical peak, 350 days post-brightening and 1500 days post-peak, as compared to the SED of the host galaxy. At peak, the SED is reasonably well fit by a $\lambda L_\lambda \propto \lambda^{-\alpha}$ power-law, with a spectral index $ \alpha=0.44\pm0.04$. A blackbody model is a very poor fit to the observed SED. At 350 days post-peak, we find $ \alpha=-0.29\pm0.05$, although we have a much more limited wavelength baseline at this epoch. Thus, over the course of the first year, the SED becomes significantly redder. By integrating the observed SED at both epochs from $0.2-4.6~\mu m$, we estimate the partial bolometric luminosities to be $(8.9 \pm 0.3) \times 10^{44}\rm ~erg~s^{-1}$ at peak and $(4.9 \pm 0.5) \times 10^{44}\rm ~erg~s^{-1}$, 350 days post-peak. Those correspond to $\simeq 24\%$ and $\simeq 13\%$ of the Eddington luminosity, for the estimated black hole mass of $10^{7.6}~M_\odot $. 

The SED at 1500 days post-peak remains nearly three times the host-galaxy contribution, indicating continued nuclear activity. To quantify the long-term fading, we extrapolated the power-law SED fitted at 350 days (black dotted line in Fig~\ref{fig:SED_fit}) into the UV. The measured UV flux at 1500 days is roughly a factor of three fainter than this extrapolated value, so the overall UV luminosity has likely declined compared to the UV luminosity at 350 days. This sustained emission may trace ongoing, low-level accretion or delayed reprocessing by circumnuclear material.
Meanwhile, the near-IR flux at this late epoch is the same as that of the host galaxy.  We also compare the UV luminosity to that of the X-ray luminosity of the source, adopting the $3\sigma$ upper limits on $L_X$ (see Section~\ref{sec:photometry}). Near peak, the limit is interestingly low with $L_X/L_{\mathrm{UV}} \leq 0.02$, while at $\sim1500$ days after peak $L_X/L_{\mathrm{UV}} \leq 16.5$, which is not a useful constraint. For comparison, AGN typically show $L_X \sim 10\%$ of the bolometric luminosity \citep{Strateva_2005}. TDEs often exhibit $L_X/L_{\mathrm{UV}} \geq 1$, and jetted TDEs can have $L_X/L_{\mathrm{UV}} \gg 1$ \citep{auchettl17}.

\subsection{Optical Spectra}
\begin{figure}
    \centering
      \includegraphics[width=\columnwidth]{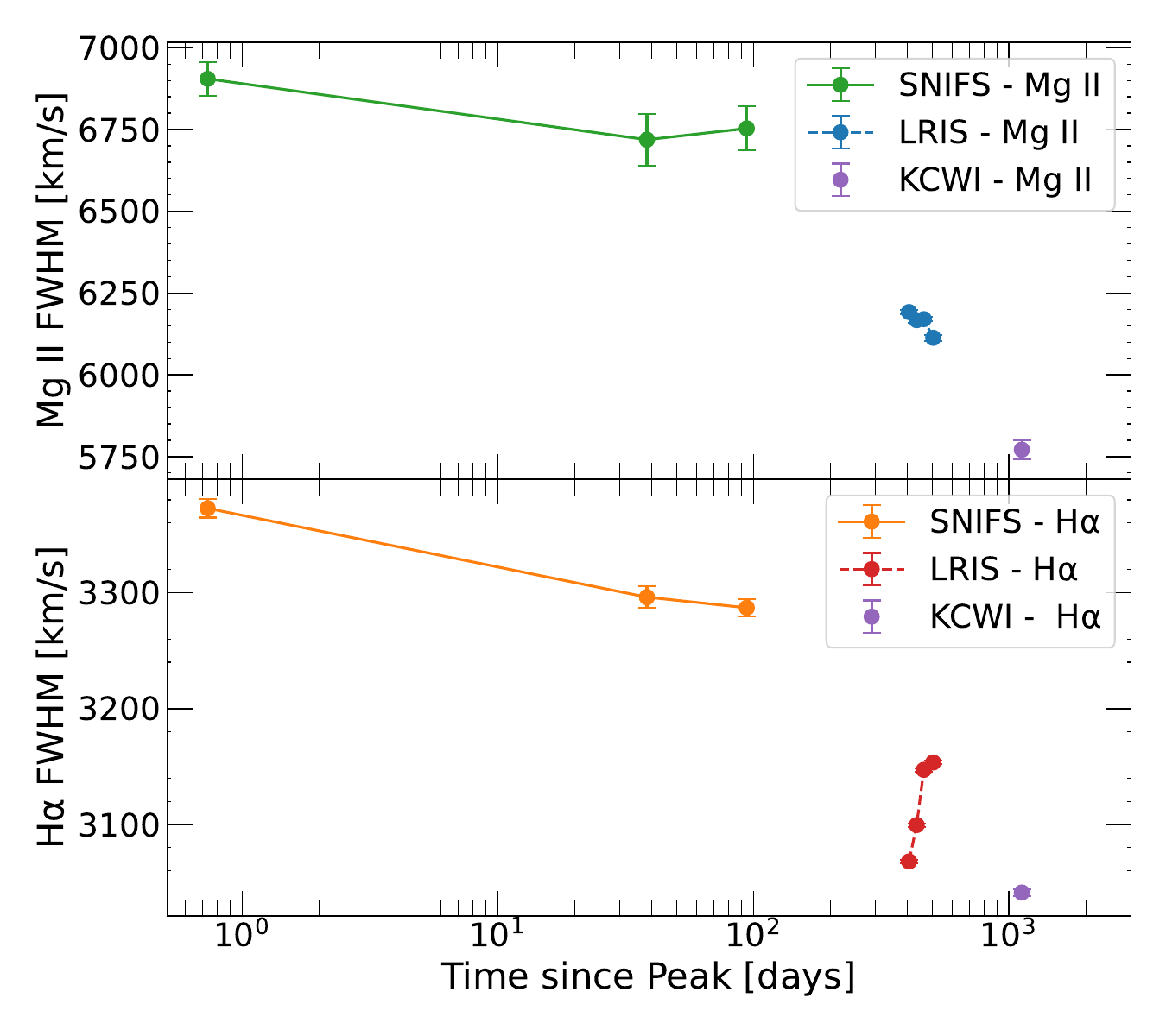}
       \caption{Evolution of the FWHM of \ion{Mg}{2} and H$\alpha$ emission lines  in the optical spectra AT2020adpi. The line width scales with the continuum flux, since as the transient fades in time, the line width decreases, which is seen in TDEs. The apparent trend in the LRIS points is just noise. } 
       \label{fig:FWHM_evolution}
   
\end{figure}

We studied the evolution of some of the strong emission lines in the optical spectra of AT2020adpi, using the {\tt astropy.specutils}  package \citep{specutils}, which estimates the FWHM of a spectral feature assuming a Gaussian profile. Figure~\ref{fig:FWHM_evolution} shows the temporal evolution of the \ion{Mg}{2} and H$\alpha$ FWHM measurements for the SNIFS, LRIS, and KCWI spectra. The SNIFS data, taken at earlier epochs (1–100 days post-peak), exhibit a clear decline in line width over time as the continuum flux decreases. The LRIS data points, observed at intermediate epochs (200–300 days), appear to deviate slightly from this trend. However, this discrepancy is likely not significant, as the statistical uncertainties on the LRIS measurements are large ($\approx 250\rm ~km~s^{-1}$). The final KCWI measurement at $\sim$1000 days also supports the trend of continued line narrowing at late times. The trend of emission lines narrowing as the transient fades is a characteristic of TDEs \citep{holoien16a, Leloudas_2019}.

The optical spectra also show an Fe~II emission complex (highlighted by green bands in Fig~\ref{fig:kcwi}). We note the presence of coronal Fe lines [FeX] 6374 and [FeVII].
These lines are commonly observed in AGNs, where they arise from high-density, partially ionized gas irradiated by a strong UV/X-ray continuum \citep[e.g.,][]{Baldwin_2004, Dong_2011}. In a typical AGN, the X-ray luminosity corresponds to roughly $10^{-2}-10^{-4}$ of the Eddington luminosity, depending on the accretion state \cite{ho08}, so, the presence of the coronal Fe lines is consistent with our upper limits on the X-ray luminosity of $L_X/L_{\rm Edd} \leq 0.003$. Coronal Fe emission has also been reported in a subset of TDEs, including AT2018fyk \citep{Wevers_2019}, ASASSN-15oi \citep{Gezari2017}, and PTF-09ge \citep{arcavi14}. \cite{Hinkle_2024_CLE} shows that many galaxies that exhibit strong coronal line emission but lack typical AGN activity have similar properties to the TDE host galaxies. Another interesting feature in the optical spectra is the presence of O III $\lambda$3132, $\lambda$3444, and N III $\lambda$4640 emission, indicating Bowen fluorescence. This requires a hard ionizing continuum producing photons above $\sim 50$ eV. 

While the SED shows a significant decline in the UV luminosity after 350 days, the spectra (Fig~\ref{fig:kcwi} and \ref{fig:spex}) obtained around this epoch indicate that substantial UV radiation is still present due to the presence of the broad H and He emission lines. In fact, if the observed \ion{He}{1} emission in the IR spectra is accurate, then its strength implies that the UV spectrum has actually hardened. The H$\alpha$ luminosity at peak from the SNIFS spectrum (refer to Figure~\ref{fig:kcwi}) is $L_{ \rm H\alpha}=5.1\times10^{41}\rm ~erg~s^{-1}$. Assuming Case B recombination and a mean ionizing photon energy of 18 eV, we estimate an ionizing luminosity of $L_{\rm ion}\simeq10^{43}\rm ~erg~s^{-1}$. We can also use the  $\mathrm{H}\alpha/\mathrm{H}\beta$ flux ratio (the Balmer decrement) to calculate the line-of-sight extinction. From our measured $\mathrm{H}\beta$ luminosity of $L_{\rm H\beta}=1.3\times10^{41}\,
\mathrm{erg\,s^{-1}}$, we obtain $\mathrm{H}\alpha/\mathrm{H}\beta\simeq 3.8$, which is larger than the intrinsic Case~B value of $2.86$ \citep{Draine2011}. Assuming the line ratio is dominated by recombination, the higher observed ratio implies a color excess of  $E(B-V)\approx 0.3$, with roughly $60\%$ of the $V$-band flux absorbed along the line of sight. However, we note that physical conditions in the BLR are far more complex than those assumed in standard Case B recombination (e.g., see the review by \citealt{Ferland_2003}). Therefore, the derived color excess should be regarded as an approximate indicator of the dust content.


\vspace{-3mm}
\subsection{Light Curve Analysis}\label{LC}
In this section, we constrain various properties of the optical and IR lightcurves, including the time of optical peak, rise time, and the lag between optical and IR lightcurves.

\subsubsection{Finding Peak and Rise Time}

We fit the ZTF $g$ and Gaia $G$ band light curves to try to estimate the time of optical peak. We first modeled the light curve using an exponential function:
\begin{equation}
m(t) = 
\begin{cases}
m_0 - A \cdot \exp\left( \dfrac{t - t_{\text{peak}}}{\tau_{\text{rise}}} \right), & \text{for } t < t_{\text{peak}} \\
m_0 - A \cdot \exp\left( -\dfrac{t - t_{\text{peak}}}{\tau_{\text{decay}}} \right), & \text{for } t \geq t_{\text{peak}}
\end{cases}
\end{equation}
where \( m_0 \) is the baseline magnitude, and \( A \) is the amplitude,  \( \tau_{\text{rise}} \) / \( \tau_{\text{decay}} \) are the rise/decay timescales. This model provides for the asymmetric shape of the light curve.
Using this model, we fit \( t_{\text{peak}} = \text{MJD}~ 59223.58 \pm 1.09\), 
\( \tau_{\text{rise}} = 27.45 \pm 0.98 \) days and  \( \tau_{\text{decay}} = 438.55 \pm 8.41\) days for the Gaia $G$ band. For the ZTF $g$ band, the values were  \( t_{\text{peak}} = \text{MJD}~ 59220.42 \pm 0.59\), 
\( \tau_{\text{rise}} = 21.42 \pm 0.56\) days and  \( \tau_{\text{decay}} = 269.10 \pm 3.67\) days. All values are in the observer frame. 
We also simply fit a parabola to the ZTF $g$ band. Using this procedure, we find \( t_{\text{peak}} = \text{MJD} ~59393.87 \pm 1.56 \). Since \( t_{\text{peak}} \) is model dependent, we simply adopt the epoch of the brightest ZTF $g$ band epoch, MJD 59378.267, as our reference time.

Next, we fit the rise time of the $g$-band light curve as a power law with
\begin{equation}
L =
\begin{cases}
k, & \text{for } t < t_{1} ~ \rm and\\
 k + h\bigg(\frac{t - t_{1}}{t_{\rm mid} -t_1}\bigg)^\alpha, & \text{for } t \geq t_{1}
\end{cases}
\end{equation}
\noindent This model fits for the baseline luminosity $k$, the time of first-light $t_1$, a luminosity scale $h$, and the power-law index $\alpha$. The denominator ($t_{\rm mid} -t_1$) normalizes the power-law term so that at $t=t_{\rm mid}$, the luminosity is $k+h$. To reduce the number of model parameters and estimate a first-order value of $\alpha$, we fix $k=1.38\times 10^{43}~\rm erg~s^{-1}$, the average luminosity measured over the year preceding the peak. We used the {\tt scipy.optimize.curvefit} package to fit the model parameters.  We get best-fit parameters of $t_1$ = MJD $59164.77\pm 1.44$, $h$ = $(5.75 \pm 0.04)  \times 10^{43}~\rm erg~s^{-1}$ and  $\alpha =  1.86 \pm 0.07$ for the rise of AT2020adpi. The value of rise-time slope is similar to the TDEs ASASSN-19bt \citep{holoien19b}, ASASSN-19dj \citep{hinkle21a}, and ZTF19abzrhgq \citep[AT2019qiz, ][]{nicholl20}, which all have a quadratic rise ($\alpha \simeq 2)$ that corresponds to a ``fireball" model with a photosphere expanding at constant velocity and temperature. 

Based on the inferred time of first light taken from the power-law fit and the time of peak light measured from the light curve in Fig~\ref{fig:all_lc}, we constrain the rise time to be $\approx170$ days in the rest frame. The rise time is considerably larger than that observed for the TDEs ASASSN-19bt \citep[$\sim 41$ days;][]{holoien19c},  PS18kh \citep[$\sim 56$ days;][]{holoien19b}, ASASSN-18pg \citep[$\sim 54$ days;][]{holoien20}, ASASSN-23bd \citep[$< 15$ days;][]{Hoogendam_2024} and ASASSN-19dj \citep[$\sim 16$ days;][]{hinkle21a}. However, the rise time of AT2020adpi is comparable to some of the ANTs in the \cite{Wiseman25} sample, such as AT2021lwx (198 days), AT2019kn (125 days), and AT2019brs (173 days).

\begin{figure*}
    \begin{center}
      \includegraphics[width=18cm]{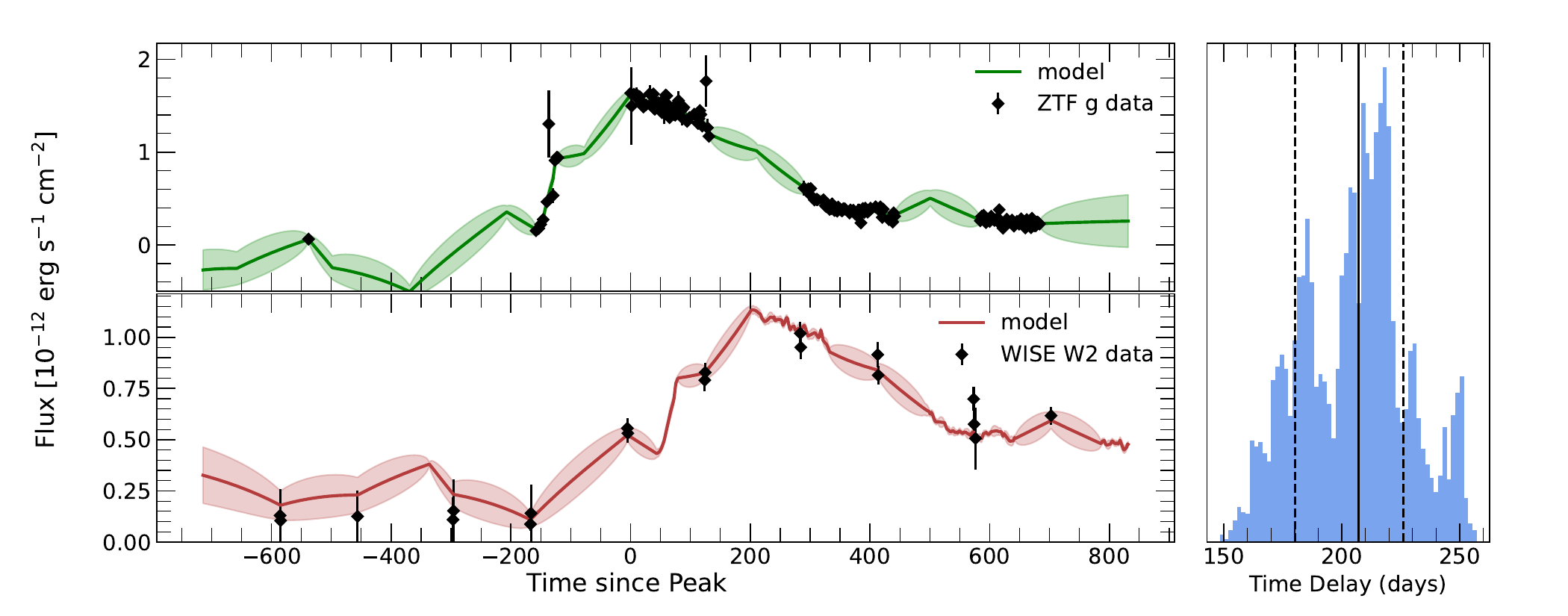}
       \caption{Left: JAVELIN model fits to the ZTF $g$ and WISE $W2$ lightcurves in rest frame, showing a best fit lag of $\approx$ 210 days in rest frame. Right: The posterior distribution for the lag from the JAVELIN models.   } 
       \label{fig:IR_lc}  
       \end{center}
\end{figure*}

\subsubsection{Light Curve Lag Analysis with JAVELIN}

In Figure~\ref{fig:all_lc}, we see that the mid-IR light curve lags the optical light curve. In order to quantify the lag, we used JAVELIN \citep{Zhu2011} to fit the ZTF $g$-band and the WISE $W2$ light curves. JAVELIN models the continuum variability using a damped random walk (DRW) Gaussian process for interpolation, and models the second light curve (W2) as a shifted, scaled, and smoothed version of the first ($g$). In this context, the DRW model is simply a sensible means of interpolating sparse, irregularly sampled light curves and does not need to have any physical meaning. Because the light curves are so sparse, we fixed the DRW damping timescale to $\tau_{\rm DRW}$ = 200 days \citep[the results are not dependent on this assumption] {McKochanek2010} and allowed only the amplitude $\sigma_{\rm DRW}$ to vary. Similarly, we fit only the lag time ($t_{\rm lag}$) and the relative amplitude, assuming no smoothing. The light curves are too sparse to allow estimates of $\tau_{\rm DRW}$ or to allow for an additional smoothing parameter. The best-fit lag value is $t_{\rm lag} \approx 208.02^{+19.37}_{-27.05}$  days in the rest frame, which corresponds to  $R = c\times t_{\rm lag} \simeq 0.2$~pc. Figure~\ref{fig:IR_lc} shows the JAVELIN modeling results for the optical (ZTF g) and mid-infrared (WISE W2) light curves.

The peak value of the optical luminosity (ZTF $g$) is 
$L_\text{peak}^{\rm opt} \simeq (3.6 \pm 0.6)\times10^{44}\,\rm erg\,s^{-1}$, and for the MIR (WISE W2), it is 
$L_\text{peak}^{\rm MIR} \simeq (2.3 \pm 0.05)\times10^{44}\,\rm erg\,s^{-1}$, making the luminosity ratio $L_{\rm MIR}/L_{\rm opt} \sim 0.6$. To order of magnitude, the luminosity of a dust echo from a dusty shell of radius $R\simeq c \times t_{\rm lag}$ is \citep{Peterson_AGN}
\begin{equation}
    L_{\rm peak}^{\rm dust} \sim \min(1, \tau)\, f\, \frac{L_{\rm peak}^{\rm opt}\, t_{\mathrm{peak}}}{t_{\mathrm{lag}}},
\end{equation}
where $t_{\mathrm{peak}}$ is the duration over which the transient remains near peak luminosity, $t_{\mathrm{lag}}$ is the time delay of the dust echo, $f$ is the covering fraction of the dust surrounding the source, and $\tau$ is the optical depth of the dust. Given that $L_{\rm peak}^{\rm dust} \sim L_{\rm peak}^{\rm opt}\, t_{\mathrm{peak}}/t_{\mathrm{lag}}$, this implies
\begin{equation}
    \min(1, \tau) \, f \sim 1.
\end{equation}
This appears to require the transient to be surrounded by dust with a high covering fraction and an optical/UV depth near or above unity. This is somewhat unexpected, as the transient exhibits a blue continuum, and the Balmer decrement suggests only $\sim 60\%$ obscuration along the line of sight. The SED also rises strongly into the UV, further suggesting low extinction in the direct viewing path beyond the Galactic contribution. This suggests that there is a considerably more UV emission at shorter wavelengths than observed by Swift, so that the pseudo-bolometric luminosity from Section~\ref{sec:spectral_energy_dist} we used for $L_{\rm opt}$ is a significant underestimate of the bolometric luminosity.

\begin{figure*}
    \begin{center}
\includegraphics[width=16cm, height=9cm]{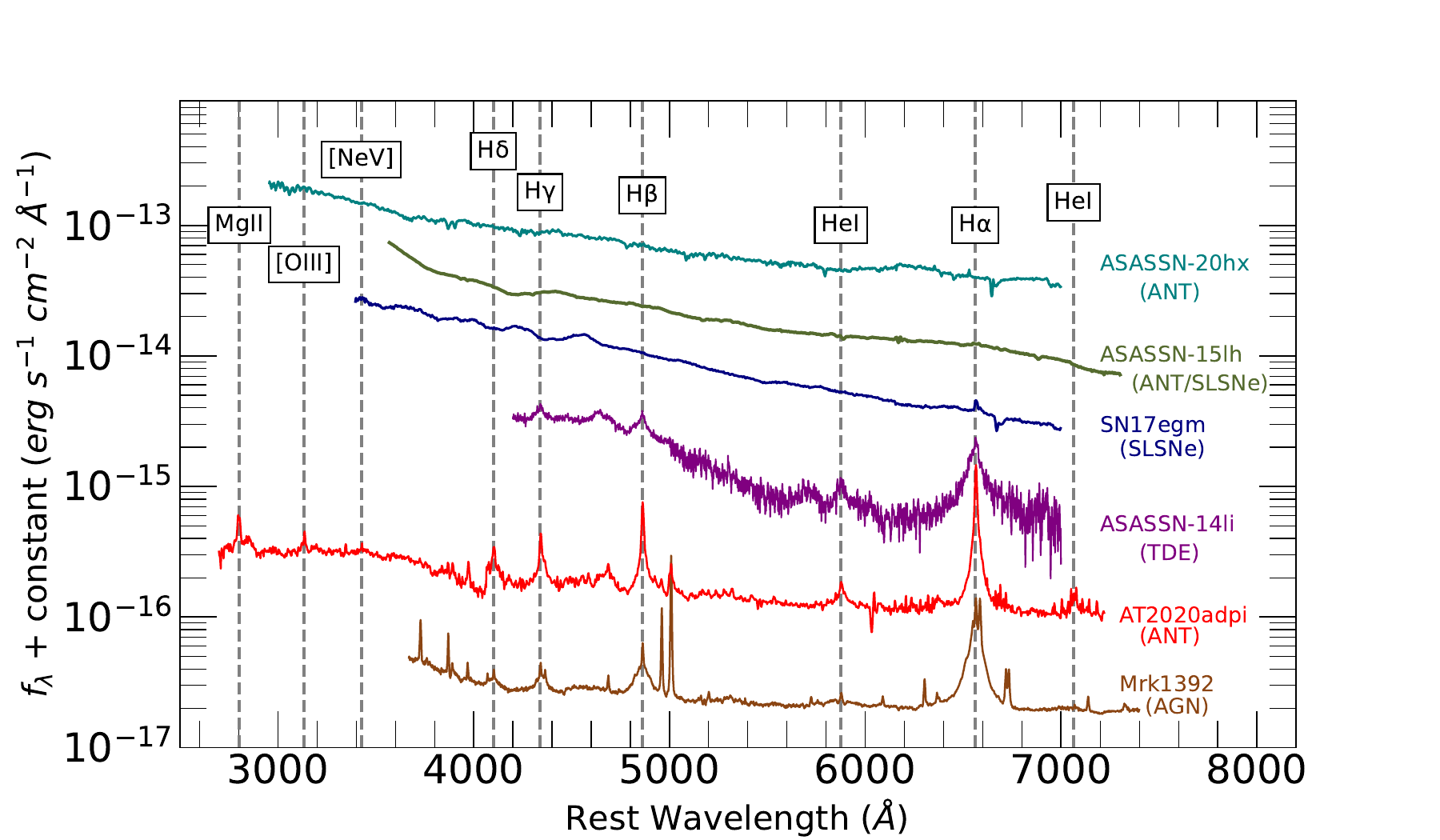}
       \caption{ The spectrum of AT2020adpi alongside those of ANT/superluminous supernova ASASSN-15lh \citep{dong16-15lh, leloudas16, godoy-rivera17}, the Type I superluminous supernova SN17egm \citep{bose18c}, the TDE ASASSN-14li \citep{holoien16-14li}, the ANT ASASSN-20hx \citep{Hinkle_2022}, and the Seyfert 1 galaxy Mrk1392 \citep{adelmanmccarthy06}. All transient spectra are shown near their peak brightness. Vertical lines indicate prominent spectral features in ANTs. The spectra are offset for visual clarity. 
} 
       \label{fig:Spactrs_comparison}
    \end{center}
\end{figure*}


\section{Discussion and Conclusions}\label{discussion}

AT2020adpi is an unusual nuclear transient with properties spanning those of both optically discovered TDEs and AGN. In Table~\ref{tab:agn_tde_comp} we summarize the TDE-like and AGN-like features. Its total radiated energy of $(1.3 \pm 0.2) \times 10^{52}$~erg is atleast an order of magnitude higher than typical supernovae but typical for luminous TDEs and bright AGN flares. There is no evidence that it is a CL-AGN or CL-LINER, as the prominent broad emission lines (Balmer lines, \ion{Mg}{2}, \ion{He}{1}, etc) are present throughout the $\sim1100$ days span of the optical spectra (see Fig~\ref{fig:kcwi}). 

\citet{Wiseman25} included AT2020adpi in a sample of 11 ambiguous nuclear transients (ANTs), analyzing ZTF, ATLAS, and Pan-STARRS optical data and NEOWISE MIR light curves. They found a $\sim$1 year lag between the optical and MIR peaks, consistent with our $\sim$240 day rest-frame lag, and similarly identified strong Balmer emission and AGN-like features. Their estimated rise time of $\sim$120 days is shorter than our $\sim$166 days, likely due to observational gaps and different peak definitions. They modeled the SED with a blackbody, finding $\log(L_{\mathrm{BB,max}}) = 44.6 \pm 0.2$~erg~s$^{-1}$ and dust temperatures of 1500–1800 K. They concluded that AT2020adpi shares traits (smooth light curve, MIR echo, strong Balmer lines) with other ANTs, possibly representing an overlap between AGN variability and massive-star TDEs.

\begin{deluxetable}{cc}[b]
\tablewidth{240pt}
\tabletypesize{\footnotesize}
\tablecaption{AGN vs. TDE Properties of AT2020adpi \label{tab:agn_tde_comp}}
\tablehead{
\colhead{Property} & 
\colhead{AGN/TDE-like}
}
\startdata
$\log(\mathrm{M}_{\mathrm{BH}}/\mathrm{M}_\odot) < 8$ & TDE/AGN \\
\ion{Fe}{2} Emission line & AGN \\
Broad \ion{Mg}{2} emission & AGN \\
$W1 - W2 > 0.7$ mag & AGN \\
Broad lines narrow as transient fades & TDE \\
Power-Law SED & AGN 
\enddata
\tablecomments{Properties used to distinguish between TDEs and AGN activity, following \cite{frederick20, Hinkle_2022}. The W1–W2 data were calculated over the entire period from MJD 55355 to MJD 60265, covering the time before, during, and after the flare event.}
\end{deluxetable}

\subsection{AT2020adpi as a TDE}

Several features initially suggested a TDE origin. The early-time optical/UV emission was bright and blue, similar to events like ASASSN-14li \citep{holoien16a} and AT2019qiz \citep{hung21}, and the optical light curve evolved smoothly without the short-timescale stochastic variability typical of AGNs \citep{hinkle21b}. The $\sim$170 day rise is slower than most TDEs (median $\lesssim$50 days; \citealt{hung17}), but luminous TDEs such as PS18kh \citep{holoien19-18kh} have shown extended rises.

The host galaxy is likely a post-starburst galaxy. This is, however, based only on the stellar age estimate from the SED models, and not a direct spectroscopic observation. TDEs frequently prefer post-starburst host galaxies \citep{french16}. No prior large-amplitude variability is detected in the available archival light curves, consistent with most TDE hosts, although there may have been a small rise in flux $\sim2$ years before the transient. The SMBH mass, $M_{\rm BH} \approx 10^{7.6}~M_\odot$, is high for TDEs \citep{wevers19}, because full disruptions of main-sequence stars become rare at these masses \citep{Hills1975,rees88}, but partial disruptions or giant-star TDEs remain possible.

The spectra, however, deviate from typical optically selected TDEs. Broad \ion{Mg}{2} $\lambda$2798 and prominent Fe II complexes are absent in most TDEs \citep{brown18, Yao2023}, but present here (Figures~\ref{fig:kcwi}–\ref{fig:spex}). 
Figure~\ref{fig:Spactrs_comparison} compares AT2020adpi's spectra with those of well-characterized TDEs and other nuclear transients. The optical spectra are similar to PS16dtm \citep{blanchard17-16dtm}, a TDE candidate in a Narrow-line Seyfert 1, which also exhibited transient blue continuum and Fe II emission, but with a different light-curve morphology. As in many TDEs, the Balmer lines narrow as the transient fades \citep{holoien19a}, opposite to the trend observed in AGNs \citep{peterson04}.

\subsection{AT2020adpi as an AGN}

Several properties point toward an AGN origin. The peak SED is well fit by a power law with $\alpha = 0.44 \pm 0.04$, flatter than most quasars but within the range for Seyfert 1 AGNs \citep{temple2023}, and inconsistent with the single-temperature blackbody typical of TDEs. The strong broad \ion{Mg}{2} and narrow Fe II emission near H$\beta$ and [O III] are typical characteristics of Narrow-Line Seyfert 1 galaxies \citep{Pogge1985, Wang_2009}.

The MIR echo peaks $\sim$210 days after the optical, corresponding to a dust radius of $\sim$0.2 pc, which is larger than typical TDE dust echoes \citep{vanvelzen16} but typical of AGN dust ``tori'' \citep{L_pez_Gonzaga_2016}. Similarly long lags have also been reported in other ANTs \citep{Wiseman25}. The high luminosity and a broad line region-like spectrum support the idea of a large-amplitude AGN flare.

However, typical AGN variability rarely produces such large, rapid changes. \citet{macleod12} found that $|\Delta m_g| > 1$ mag in $\lesssim$150 days occurs with a probability $\lesssim 10^{-5}$ in quasars. CL-AGNs can reach these amplitudes over months \citep{shappee16, macleod16}, but usually show spectral evolution toward a normal AGN continuum and line ratios. Due to a lack of any pre-flare optical spectra, we cannot determine if AT2020adpi is associated with a CL-AGN.

\begin{figure}
    \begin{center}
\includegraphics[width=9cm, height=9cm]{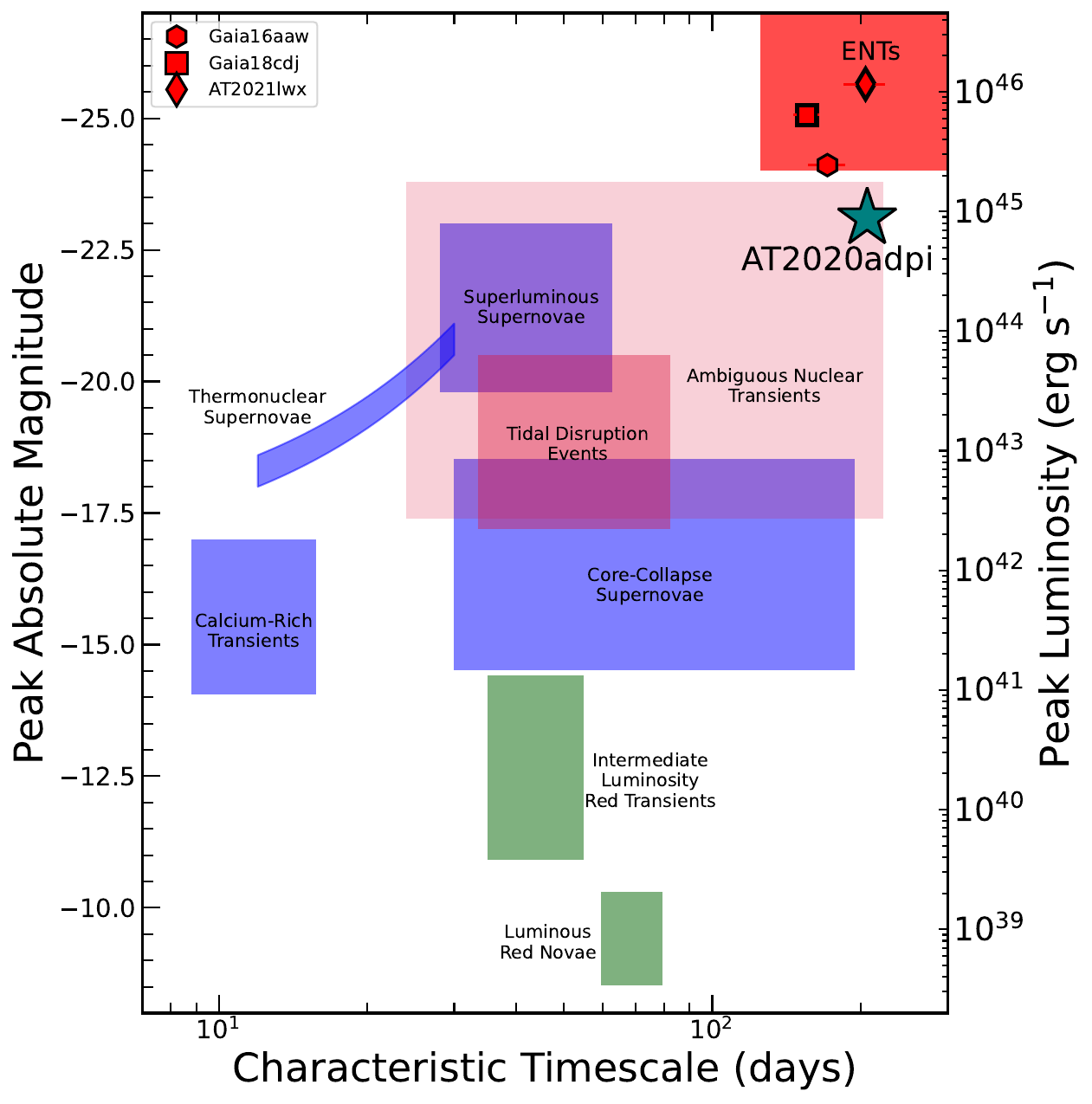}
       \caption{Comparison of AT2020adpi with other known transients in the optical absolute magnitude versus characteristic timescale parameter space, adapted from \citet{Hinkle_science_25}. The blue-shaded regions represent various classes of SNe, green regions correspond to stellar mergers and/or mass transfer events, and red-shaded regions indicate transients powered by accretion onto SMBHs. The sample of ENTs described in \cite{Hinkle_science_25} is marked in red symbols. AT2020adpi shares features similar to ANTs and ENTs.} 
       \label{fig:all_transients}
    \end{center}
\end{figure}

\subsection{AT2020adpi as an ANT}

The combination of a smooth, luminous light curve, Balmer line narrowing, post-starburst host galaxy, power-law continuum, \ion{Mg}{2} and Fe II emission, and a torus-scale MIR echo suggests AT2020adpi is an ANT \citep{Wiseman25, Hinkle2024}. In this scenario, a TDE-like fueling episode, which can possibly be a partial stellar disruption, occurs in a galaxy with a pre-existing AGN, producing both a TDE-like response and AGN-like continuum and line features. The high dust covering fraction we infer for AT2020adpi is typical for AGNs, near the upper end values measured for ANTs \citep[$f_c = 0.29$;][]{Hinkle_2024}, and significantly higher than the values found in optically selected TDEs. Similar ambiguous cases include PS1-10adi \citep{kankare17}, ZTF18aajupnt \citep{frederick19}, ASASSN-18jd \citep{neustadt20}, and ASASSN-20hx \citep{hinkle22}, all of which show mixed TDE/AGN traits.

Figure~\ref{fig:all_transients} places AT2020adpi within the broader distribution of nuclear transients, based on the typical peak absolute magnitude and the time it takes to fade to half of its peak luminosity of each class. Its location in the upper-right region of this diagram corresponds to events with high total radiated energy. In this context, many of AT2020adpi’s properties align with those seen in other ANTs. The rise time is also consistent with those of the ANTs \citep{Wiseman25}.

The increasing sample of ANTs highlights the challenge of classification in the era of wide-field time-domain surveys and enhances the need for multi-wavelength, spectroscopic, and long-term monitoring to disentangle their physical origins. We anticipate finding more such events with the Rubin LSST survey, which will help us decipher their light curves, spectra, and host galaxy properties.

\section*{Acknowledgements}
We thank the Swift observatory team for promptly scheduling and executing our TOO observations. PP thanks Rick Pogge for valuable comments on the KCWI spectrum and the Transients from Space Workshop participants at STScI for helpful suggestions and discussions. CSK is supported by NSF grants AST-2307385 and AST-2407206. Parts of this research were supported by the Australian Research Council Discovery Early Career Researcher Award (DECRA) through project number DE230101069.

\newpage

\bibliographystyle{mnras}
\bibliography{biblio} 

\end{document}